# P-doped organic semiconductor: potential replacement for PEDOT:PSS in organic photodetectors

J. Herrbach[1], A. Revaux[1,a)], D. Vuillaume[2], and A. Kahn[3]

[1]Univ. Grenoble Alpes, CEA-LITEN, Grenoble, 38000, France

[2]IEMN, CNRS, Univ. Lille, Villeneuve d'Ascq, 59652, France

[3]Dept. of Electrical Engineering, Princeton University, Princeton, NJ, 08544, USA

In this work we present an alternative to the use of PEDOT:PSS as hole transport and electron blocking layer in organic photodetectors processed by solution. As PEDOT:PSS is known to be sensitive to humidity, oxygen and UV, removing this layer is essential for lifetime improvements. As a first step to achieving this goal, we need to find an alternative layer that fulfills the same role in order to obtain a working diode with similar or better performance. As a replacement, a layer of Poly[(4,8-bis-(2-ethylhexyloxy)-benzo(1,2-b:4,5-b′)dithiophene)-2,6-diyl-alt-(4-(2-ethylhexanoyl)-thieno[3,4-b]thiophene-)-2-6-diyl)] (PBDTTT-c) p-doped with the dopant tris-[1-(trifluoroethanoyl)-2-(trifluoromethyl)ethane-1,2-dithiolene] (Mo(tfd-COCF$_3$)$_3$) is used. This p-doped layer effectively lowers the hole injection barrier, and the low electron affinity of the polymer prevents the injection of electrons in the active layer. We show similar device performance under light and the improvements of detection performance with the doped layer in comparison with PEDOT:PSS, leading to a detectivity of $1.9 \times 10^{13}$ cm(Hz)$^{1/2}$(W)$^{-1}$, competitive with silicon diodes used in imaging applications. Moreover, contrary to PEDOT:PSS, no localization of the p-doped layer is needed, leading to a diode active area defined by the patterned electrodes.

Photodetectors based on organic active layers have shown significant advantages over the past decade. The use of polymers instead of inorganic materials like silicon or germanium leads to a strong reduction in processing cost and opens possibilities for applications on large area, flexible and transparent substrates [1–4]. Today, organic devices exhibit good electrical performance and can compete with amorphous silicon, but improvements need to be made to increase their lifetime.

a) Author to whom correspondence should be addressed. Electronic mail: amelie.revaux@cea.fr.

Most organic photodetectors (OPDs) are made with a blend of a polymer donor and fullerene acceptor molecules as active layer. Charge transport and blocking layers are used to adapt the work function of the electrodes and reduce dark current densities in the reverse bias regime [2]. PEDOT:PSS is widely used as hole transport layer (HTL) and electron blocking layer (EBL) [5]. However, this conductive polymer is known to be sensitive to humidity, oxygen and UV, leading to rapid aging [6–9]. Attempts to eliminate PEDOT:PSS from the OPDs have already been carried out, leading to improvements in device lifetime [10]. However, the suppression of the HTL-EBL also degrades the dark current density at high reverse bias.

As in inorganic devices, doping the semiconductor at the interface is an alternative to creating efficient contacts, and in the present case, to the use of PEDOT:PSS. The introduction of a thin p-doped semiconductor layer between the electrode and the active blend yields good injection performance. Carriers tunnel through the narrow depletion width effectively lowering the injection barrier [11]. With a properly chosen semiconductor with low electron affinity (EA) or high lowest unoccupied molecular orbital (LUMO), the electron blocking role of PEDOT:PSS is also maintained. Doping the semiconductor also leads to high conductivity, enhancing charge transport in these interface layers [12]. P- and n-doped organic semiconductors obtained by co-evaporation are widely used for OLEDs where efficient charge injection and low resistance decrease the operating voltage and increase the power efficiency of the device [13–16]. In printed electronics, the processability of multiple organic layers is limited and requires the development of new techniques, such as lamination processes [17,18].

This work focuses on the analysis of an OPD based on a blend between the donor polymer Poly[(4,8-bis-(2-ethylhexyloxy)-benzo(1,2-b-4,5-b')dithiophene)-2,6-diyl-alt-(4-(2-ethylhexanoyl)-thieno[3,4-b]thiophene-)-2-6-diyl)] (PBDTTT-c) and the acceptor [6,6]-phenyl-C61-butyric acid methyl ester (C60-PCBM). The HTL-EBL is introduced by soft contact transfer lamination (SCTL) [19] of a thin PBDTTT-c layer p-doped with the complex molybdenum tris-[1-(trifluoroethanoyl)-2-(trifluoromethyl)ethane-1,2-dithiolene] (Mo(tfd-COCF$_3$)$_3$). This derivative of molybdenum tris-[1,2-bis(triuoromethyl)ethane-1,2-dithiolene] (Mo(tfd)$_3$) [20] shows good doping efficiency and high solubility in organic solvents [21]. Moreover, this dopant has been carefully chosen for its 3D structure, which reduces its diffusion in organic layers, in particular PBDTTT-c [22]. The contact between the HTL-EBL and the active layer is achieved via SCTL.

The present work suggests an alternative to PEDOT:PSS as HTL-EBL in organic photodetectors with the integration of the doped polymer layer leading to similar performance under light and improvement of the detectivity.

In this study, devices with two different HTL-EBL are compared and all devices consist of the same structure: Glass / ITO-PEIE (110 nm) / Blend (500±20 nm) / HTL-EBL / Aluminum (100 nm). The stack and energy levels are shown on Figure 1. ITO is patterned on ITO-coated glass substrates by photolithography and recovered by a thin layer of Polyethylenimine ethoxylated (PEIE) deposited by spin-coating, annealed at 100°C and rinsed. The active layer is composed of PBDTTT-c and C60-PCBM with a 1:1.5 ratio in weight, spin-coated on ITO/PEIE and annealed at 115°C to form a 500 nm layer. In device A, PEDOT:PSS is used as HTL-EBL with a thickness of 180±10 nm. To obtain a good comparison between both devices, PEDOT:PSS is deposited by SCTL on the active layer following the technique described by Gupta *et al.* [23]. In Device B, the HTL-EBL is changed to layer of PBDTTT-c doped with Mo(tfd-COCF$_3$)$_3$ with a thickness of 45±10 nm. The dopant was synthesized by the group of Prof. Marder at the Georgia Institute of Technology. With an electron affinity around 5.3 eV, it has been shown to efficiently p-dope PBDTTT-c [22]. In this work we use a PBDTTT-c:Mo(tfd-COCF$_3$)$_3$ blend with a 5% molar ratio. Since the doped polymer cannot be directly spin-coated on top of the active layer without dissolving the previous layer, SCTL is used as described elsewhere [19]. For both devices A and B, an aluminum top electrode is evaporated through a shadow mask in a vacuum chamber. The devices are then encapsulated in the glovebox.

A second type of structure was processed to study the influence of the doping concentration on the hole injection barrier (see inset of Figure 2). This device is composed of Glass / ITO (110 nm) / PEDOT:PSS (40±10 nm) / Blend (90±10 nm) / interface layer (45±10 nm). The PEDOT:PSS layer is spin-coated on ITO-coated glass substrates. The active layer composed of PBDTTT-c:C60-PCBM is spin-coated on PEDOT:PSS and the interface layer is deposited by SCTL. For this study, the pure polymer (PBDTTT-c) and 3 doping concentrations (1, 1.5 and 5%) are tested as interface layer. This device was then characterized using a controlled growth mercury electrode as top contact. The diameter of the circular contact is determined with a 10% error using a lateral webcam.

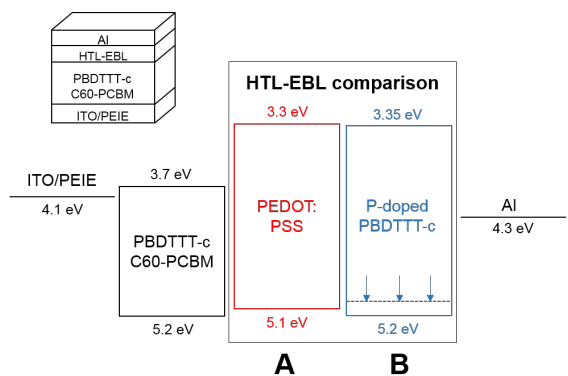

FIG. 1. Energy level diagram of the different components. The inset shows the stack of layers included in the photodetector studied here.

The effect of doping has been studied for numerous polymer-dopant couples. Expected impacts of doping are an increase in conductivity, a decrease in hopping transport activation energy and a shift of the Fermi level (towards the HOMO of the polymer for p-doping) [24–26]. The increase in charge density also leads to an increase in carrier injection at organic/electrode interfaces, through either a reduction of the barrier or a reduction of the depletion width [27], resulting in an increase in carrier injection efficiency.

A study of hole injection into the PBDTTT-c:C60-PCBM blend as a function of doping concentration was carried out with the device shown in Figure 2. PEDOT:PSS enables the injection of holes from the ITO electrode to the blend and the doped polymer reduces the hole injection barrier from a mercury electrode. Injection from the laminated polymer layer p-doped with different concentrations (open shapes) is compared to PEDOT:PSS (solid shapes) in Figure 2.

With a pure (undoped) polymer layer between the blend and the mercury electrode, a difference of 5 orders of magnitude is measured for the injected current density at 2 V between both electrodes. The addition of dopant in the laminated polymer layer leads to a reduction of the injection barrier and an increase in the hole current density injected from the mercury electrode. With a 1.5% molar ratio, the hole injection is similar for both electrodes and with a 5% molar ratio, the injection current from mercury electrode is higher than from ITO/PEDOT:PSS.

These results confirm that p-doped PBDTTT-c is an efficient alternative to PEDOT:PSS as an HTL enabling the injection of holes in the blend. Regarding the results, a molar ratio of 5% is chosen for the HTL-EBL integrated in the photodetector.

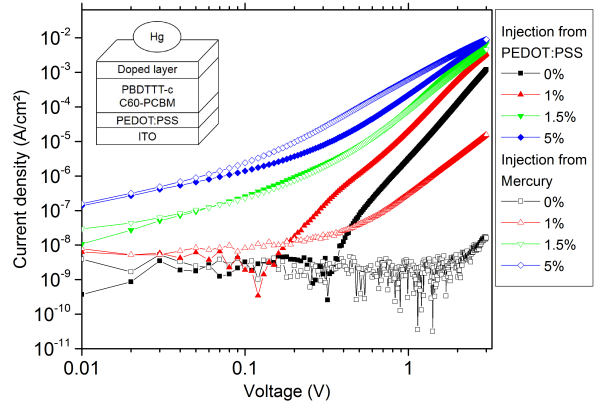

FIG. 2. Diagram of the test device structure used to study the injection barrier with doping as inset and evolution of the current density injected from PEDOT:PSS and from mercury as a function of the doping concentration of the doped layer.

Figure 3 (a) shows dark and light (530 nm, 650 mW/m²) I-V characteristics for devices A (red) and B (blue). These devices are also compared with a photodetector processed without HTL-EBL (grey).

When no HTL-EBL is added to the device (grey curve in Figure 3 (a)), the light current density in the reverse regime is similar to that in devices A and B. No injection barrier lowering or work-function adjustment is needed in this regime. This technique is used by Ramuz *et al.* [10] to improve the device lifetime by avoiding PEDOT:PSS. However, the suppression of the electron blocking function of the layer leads to a strong increase of the dark current density in the reverse regime with increasing electric field in the device. For OPD application, the dark and light current densities in the reverse regime need to be electric field-independent to allow for small voltage variations around the bias set by the electronics. To prevent this increase in dark current density, electron and hole blocking layers (EBL and HBL) are used to prevent injection of minority carriers in reverse bias regime. In this work, the HTL, i.e., PEDOT:PSS or p-doped PBDTTT-c, also acts as EBL.

The thickness of the PEDOT:PSS (180±10 nm) layer in device A was optimized under the following considerations. A decrease of the PEDOT:PSS thickness leads to better performance under illumination, but increases the dark current in the reverse bias regime (see Ref. [28]). A trade-off needs to be made between dark and light performances, since good detectivity is critical for photodetection applications, contrary to photovoltaic applications where the photon conversion and charge extraction are essential.

We notice in Figure 3 (a) the decrease of the dark current density in the reverse bias regime for device B compared to device A. At -2 V, the dark current density is reduced by one order of magnitude to reach $6.7 \times 10^{-10}$ A/cm². Performance under illumination is similar for both devices with almost identical light current densities in the reverse bias regime and external quantum efficiencies (EQE). Figure 3 (b) shows essentially identical EQE for both devices between 380 and 940 nm at -2 V, reaching 72% and 73% at 640 nm for the PEDOT:PSS and p-doped PBDTTT-c devices, respectively.

In order to compare the performances of these diodes with those reported in the literature, the detectivity is determined at -2 V and 530 nm. If the Johnson and thermal noises are neglected, as they are in most publications [2,29,30], the detectivity depends on the shot noise following the relation

$$D^* = q\lambda EQE/hc \times (2qJ_d)^{-1/2} \quad , \tag{1}$$

where $\lambda$ is the wavelength of the light and $J_d$ the dark current density. Detectivities of $5.52 \times 10^{12}$ and $1.93 \times 10^{13}$ cm(Hz)$^{1/2}$(W)$^{-1}$ are obtained for devices A and B, respectively. The improvement of the dark current density in the devices using the doped laminated layer leads to a detectivity increase of one order of magnitude compared to the PEDOT:PSS device. Moreover, this performance is among the best detectivities reported in the literature [31] and competes with those of silicon-made devices with typical detectivities around $10^{13}$ cm(Hz)$^{1/2}$(W)$^{-1}$ for imaging applications.

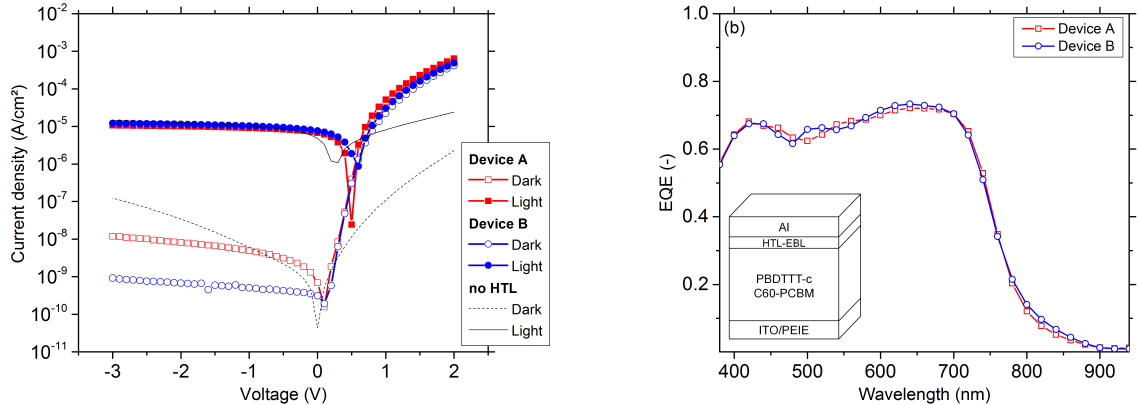

FIG. 3. Current density in the dark and under illumination (530 nm, 650 mW/m²) for PEDOT:PSS (red) and the p-doped polymer (blue) as HTL-EBL, and without HTL-EBL (grey) (a). EQE at -2V for PEDOT:PSS (red) and the the p-doped layer (blue) as HTL-EBL. The stack of the photodetector is added as inset (b).

SCTL is used here to deposit a doped polymer layer on top of the active blend. To insure that this technique does not introduce inhomogeneities on the diode, we map the sensitivity of the diode on a 16 mm² area. Figures 4 (a) and (b) represent the sensitivity scans of the diodes with PEDOT:PSS and p-doped PBDTTT-c, respectively. The sensitivity mapping contains the 3.14 mm² diode and a part of the electrical tracks in ITO and Aluminum. A top view representation of the device is given on Figure 4 (c).

We notice that the sensitivity is similar for devices A and B with 0.26 A/W, which is consistent with the EQE and light current density measurements of Figures 3 (a) and (b). No sensitivity inhomogeneities are detected on the active area containing the laminated doped layer. Moreover, the sensitivity scans highlight a strong difference between the p-doped polymer and PEDOT:PSS. Device A displays sensitivity only on the diode area defined by patterned ITO and Aluminum, whereas PEDOT:PSS can act as an electrode without the use of aluminum. In the latter case, the area of the ITO recovered by the blend and PEDOT:PSS shows a sensitivity 0.02 A/W lower than in the center of the diode. This implies that the active area is given by the superposition of ITO and PEDOT:PSS.

These different behaviors are presumably due to the 6 orders of magnitude difference between the conductivities of the two materials. The doped polymer is used to lower the effective barrier between the blend and the aluminum electrode, but its conductivity, which is around $10^{-4}$ S/cm, is insufficient to transport the extracted holes laterally towards the aluminum area where they can be collected. With a conductivity measured at 425 S/cm, PEDOT:PSS can act as a conducting track and is widely used in printed electronics as transparent electrode.

The similar results obtained for current density, EQE and sensitivity measurements show therefore that a high conductivity is not necessary to obtain a good extraction of charges. Effective barrier lowering drives the injection of charges when a thin doped layer is used as HTL-EBL. Moreover, a lower conductivity has advantages for processing. The use of a doped polymer layer as HTL does not require localization, in contrast to PEDOT:PSS, and no constraint is given for the choice of the conducting top electrode.

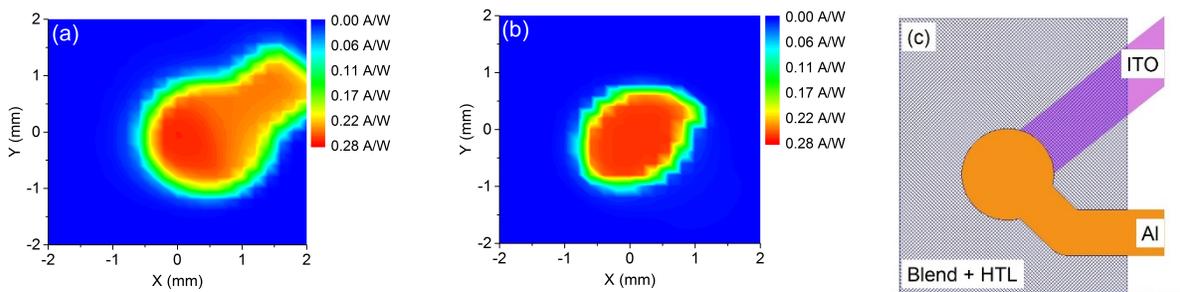

FIG. 4. Sensitivity map at 530 nm, 16.7 mW/m², for the diode with PEDOT:PSS (a) and p-doped PBDTTT-c (b) as HTL-EBL and top view representation of the device (c).

In this work, PEDOT:PSS usually used as HTL and EBL is replaced in an organic photodetector by a thin doped polymer layer deposited by soft contact transfer lamination. The performances of these two device stacks are compared with standard figures of merits for photodetectors. While keeping similar performances under illumination, the use of the doped layer enhances the detection of the photodetector over that achieved with PEDOT:PSS.

Finally, the suppression of PEDOT:PSS and its replacement with a p-doped polymer using a non-diffusive dopant is likely to lead to better stability performance of the photodetector, a topic that will be the subject of future work.

## ACKNOWLEDGMENTS

A part of this work was carried out at Princeton University and financially supported by a grant of the National Science Foundation (DMR-1506097). AK thanks the group of Prof. Seth Marder for providing the dopants.

## REFERENCES

[1] A.C. Arias, J.D. MacKenzie, I. McCulloch, J. Rivnay, and A. Salleo, Chem. Rev. **110**, 3 (2010).

[2] K.J. Baeg, M. Binda, D. Natali, M. Caironi, and Y.Y. Noh, Adv. Mater. **25**, 4267 (2013).

[3] X. Liu, H. Chen, and S. Tan, Renew. Sustain. Energy Rev. **52**, 1527 (2015).

[4] B. Xiao, H. Wu, and Y. Cao, Mater. Today **18**, 385 (2015).

[5] P. Cheng and X. Zhan, Chem. Soc. Rev. **45**, 2544 (2016).

[6] H. Cao, W. He, Y. Mao, X. Lin, K. Ishikawa, J.H. Dickerson, and W.P. Hess, J. Power Sources **264**, 168 (2014).

[7] E. Voroshazi, B. Verreet, A. Buri, R. Müller, D. Di Nuzzo, and P. Heremans, Org. Electron. Physics, Mater. Appl. **12**, 736 (2011).

[8] K. Norrman, M. V. Madsen, S. a. Gevorgyan, and F.C. Krebs, J. Am. Chem. Soc. **132**, 16883 (2010).

[9] S.B. Sapkota, M. Fischer, B. Zimmermann, and U. Würfel, Sol. Energy Mater. Sol. Cells **121**, 43 (2014).

[10] M. Ramuz, L. Bürgi, C. Winnewisser, and P. Seitz, Org. Electron. Physics, Mater. Appl. **9**, 369 (2008).

[11] W. Gao and A. Kahn, Org. Electron. Physics, Mater. Appl. **3**, 53 (2002).

[12] K.H. Yim, G.L. Whiting, C.E. Murphy, J.J.M. Halls, J.H. Burroughes, R.H. Friend, and J.S. Kim, Adv. Mater. **20**, 3319 (2008).

[13] G. He, O. Schneider, D. Qin, X. Zhou, M. Pfeiffer, and K. Leo, J. Appl. Phys. **95**, 5773 (2004).

[14] J. Huang, M. Pfeiffer, A. Werner, J. Blochwitz, K. Leo, and S. Liu, Appl. Phys. Lett. **80**, 139 (2002).

[15] J.-H. Lee and J.-J. Kim, J. Inf. Disp. **14**, 39 (2013).

[16] J.H. Lee and J.J. Kim, Phys. Status Solidi Appl. Mater. Sci. **209**, 1399 (2012).

[17] L. Chen, P. Degenaar, and D.D.C. Bradley, Adv. Mater. **20**, 1679 (2008).

[18] A.L. Shu, A. Dai, H. Wang, Y.L. Loo, and A. Kahn, Org. Electron. Physics, Mater. Appl. **14**, 149 (2013).

[19] A. Dai, Y. Zhou, A.L. Shu, S.K. Mohapatra, H. Wang, C. Fuentes-Hernandez, Y. Zhang, S. Barlow, Y.L. Loo, S.R. Marder, B. Kippelen, and A. Kahn, Adv. Funct. Mater. **24**, 2197 (2014).

[20] Y. Qi, T. Sajoto, S. Barlow, E.G. Kim, J.L. Brédas, S.R. Marder, and A. Kahn, J. Am. Chem. Soc. **131**, 12530 (2009).

[21] J. Belasco, S.K. Mohapatra, Y. Zhang, S. Barlow, S.R. Marder, and A. Kahn, Appl. Phys. Lett. **105**, 2012 (2014).

[22] A. Dai, Creating Highly Efficient Carrier Injection or Collection Contacts via Soft Contact Transfer Lamination of P-Doped Interlayers, Princeton University, 2015.

[23] D. Gupta, M.M. Wienk, and R. a J. Janssen, Adv. Energy Mater. **3**, 782 (2013).

[24] I. Salzmann and G. Heimel, J. Electron Spectros. Relat. Phenomena **204**, 1 (2015).

[25] S. Olthof, S. Mehraeen, S.K. Mohapatra, S. Barlow, V. Coropceanu, J.L. Brédas, S.R. Marder, and A. Kahn, Phys. Rev. Lett. **109**, 1 (2012).

[26] M.L. Tietze, P. Pahner, K. Schmidt, K. Leo, and B. Lüssem, Adv. Funct. Mater. **25**, 2701 (2015).

[27] S. Olthof, W. Tress, R. Meerheim, B. Lüssem, and K. Leo, J. Appl. Phys. **106**, 103711 (2009).

[28] See supplementary material at [URL will be inserted by AIP] for additional data on the impact of PEDOT:PSS thickness on IV characteristics.

[29] X. Gong, M. Tong, Y. Xia, W. Cai, J.S. Moon, Y. Cao, G. Yu, C.-L. Shieh, B. Nilsson, and A.J. Heeger, Science (80-. ). **325**, 1665 (2009).

[30] J.R. Manders, T.H. Lai, Y. An, W. Xu, J. Lee, D.Y. Kim, G. Bosman, and F. So, Adv. Funct. Mater. **24**, 7205

(2014).

[31] A. Pierre, I. Deckman, P.B. Lechêne, and A.C. Arias, Adv. Mater. **27**, 6411 (2015).